\documentclass[twocolumn,aip,reprint,superscriptaddress]{revtex4-2}

\usepackage{amsmath,amsfonts}
\usepackage{graphicx}

\usepackage{booktabs}
\usepackage{dcolumn}

\usepackage{bm}
\usepackage{physics}
\usepackage{xcolor}
\usepackage{enumitem} 
\usepackage{comment}

\newcommand{\chem} {Department of Chemistry, School of Science and Research Center for Industries of the Future, Westlake University, Hangzhou 310030, P. R. China.}
\newcommand{\study} {Institute of Natural Sciences, Westlake Institute for Advanced Study, Hangzhou 310024, China.}
\newcommand{\phys} {Department of Physics, School of Science, Westlake University, Hangzhou 310030, P. R. China.}

\begin{document}

\title{A DMFT approach to evaluate electronic frictional effects near solid surfaces of strongly correlated systems}

\author{Yunhao Liu}
\affiliation{\chem}
\affiliation{\study}

\author{Wenjie Dou}%
\email{douwenjie@westlake.edu.cn} 
\affiliation{\chem}
\affiliation{\study}
\affiliation{\phys}

\date{\today}

\begin{abstract}
Electronic friction–Langevin dynamics (EF-LD) provides an efficient framework for capturing nonadiabatic effects at solid surfaces, with particular relevance to electrochemistry and molecular electronics. In this work, we investigate electronic friction in the two-dimensional Hubbard–Holstein model employing dynamical mean-field theory (DMFT), where the full density-matrix numerical renormalization group (FDM-NRG) serves as the impurity solver. Our results are benchmarked against mean-field theory (MFT). DMFT yields two distinct peaks in the electronic friction, arising from electron attachment/detachment resonances with the solid Fermi level, whereas MFT is unable to capture this Fermi resonance. We further examine the dynamics of electronic friction via EF-LD simulations. Our simulations uncover significant discrepancies mainly in the electronic population evolution predicted by MFT versus DMFT, indicating that MFT is inadequate for describing nonadiabatic dynamics in strongly correlated systems. Thanks to its flexibility and computational efficiency, the proposed DMFT-based approach can be readily extended to a broad range of applications.
\end{abstract}

\maketitle

\section{\label{sec:sec1} Introduction}

The Born-Oppenheimer approximation breaks down in scenarios where nuclear motion interfaces strongly with a continuum of electronic states, such as in molecule-solid surface interactions. This is evident from experiments on gas-surface scattering\cite{huang2000vibrational,doi:10.1126/science.aad4972}. While numerically exact quantum methods like the hierarchical equations of motion (HEOM)\cite{doi:10.1143/JPSJ.58.101,doi:10.1143/JPSJ.75.082001,10.1063/5.0011599} and the multiconfiguration time-dependent Hartree (MCTDH)\cite{MEYER199073,10.1063/1.463007,Meyer2003} exist for systems with limited degrees of freedom, such approaches prove computationally intractable for realistic interfacial systems. To overcome this limitation, the generalized Langevin dynamics framework has emerged as a key computational paradigm for simulating nonadiabatic dynamics at the molecule-solid interfaces. In this framework, nonadiabatic dynamics is incorporated into electronic friction and random forces\cite{10.1063/1.440287,10.1063/1.432526}, thereby capturing essential electron transfer between the molecule and the solid surfaces while maintaining computational feasibility.

The electronic friction represents the first order correction to the Born-Oppenheimer approximation, serving as a fundamental mechanism for understanding nonadiabaticity in a electronic bath induced by solid surface. The concept of electronic friction has been widely applied in diverse fields, including electrochemistry \cite{doi:10.1021/jp805876e,doi:10.1021/acs.jpcc.5b06655,B805544K,doi:10.1021/jp9933673,PhysRevLett.84.1051} and molecular electronics\cite{doi:10.1021/jp7114548,Tao2006,Galperin_2007,PhysRevLett.100.176403,doi:10.1126/science.1081572,PhysRevB.88.201405,PhysRevB.76.085433}. For example, electronic friction has been used to explain the energy loss of gas molecules scattering off solid surfaces\cite{huang2000vibrational,doi:10.1021/acs.jpclett.5b02448,C0CP02086A}. Moreover, electronic friction has also been shown to be relevant for chemisorption and dissociation\cite{doi:10.1126/science.aad4972,PhysRevLett.118.256001,PhysRevLett.117.196001}. 

The formalism of electronic friction is rooted in a physical picture where nuclei interact with fast-relaxing electronic bath. Within this picture, nuclear motion perturbs the electronic steady state. Because the electronic bath relaxes much faster than the nuclei move, the delayed electronic feedback on the nuclear dynamics can be approximated as instantaneous. Consequently, this back-action is effectively modeled by a Markovian damping force supplemented by a stochastic thermal noise term that obeys the fluctuation-dissipation theorem\cite{PhysRevLett.119.046001}. The nuclear coordinates $\mathbf{R}$ thus obey a Langevin equation of the form:
\begin{equation}\label{eqn-1}
m_{\alpha} \ddot{R}_{\alpha}=\bar{F}_{\alpha}-\sum_{v} \gamma_{\alpha v} \dot{R}_{v}+\zeta_{\alpha}(t),
\end{equation}
where $\alpha$ and $\nu$ are index nuclear degrees of freedom (DoFs), $\bar{F}_\alpha = -\operatorname{tr}_e\left(\partial_\alpha\hat{H}\rho_\mathrm{ss}\right)$ represents the mean force, $\gamma_{\alpha\nu}$ the (Markovian) friction tensor, and $\zeta_{\alpha}(t)$ the random force component. In our previous work\cite{PhysRevLett.119.046001}, we have derived a universal electronic friction from a quantum-classical Liouville equation (QCLE), which should be valid in and out of the electronic equilibrium, with or without electron-electron (el-el) interactions:
\begin{equation}\label{eqn-2}
\gamma_{\alpha \nu}=-\int_{0}^{\infty} \mathrm{d} t \operatorname{tr}_{e} \left(\partial_{\alpha} \hat{H} \mathrm{e}^{-\mathrm{i} \hat{H} t / \hbar} \partial_{\nu} \hat{\rho}_\mathrm{ss} \mathrm{e}^{\mathrm{i} \hat{H} t / \hbar}\right),
\end{equation}
where $\hat{H}\left(\mathbf{R}\right)$ is the electronic Hamiltonian, $\hat{\rho}_\text{ss}\left(\mathbf{R}\right)$ is the steady state's electronic density matrix, and $\mathrm{tr}_e$ implies tracing over many-body electronic states. At equilibrium, the steady-state electronic density matrix is $\rho_\text{ss}=\mathrm{e}^{-\hat{H}/k_\text{B}T}/Z$, where $Z\equiv\operatorname{tr}_e\left(\mathrm{e}^{-\hat{H}/k_\text{B}T}\right)$ is the corresponding partition function. Based on this, we have proved the second fluctuation-dissipation theorem\cite{PhysRevLett.119.046001} is satisfied at equilibrium, thereby the friction tensor can be represented as
\begin{equation}\label{eqn-3}
    \gamma_{\alpha \nu} =\frac{\beta}{2} \int_{0}^{\infty} \mathrm{d}t \operatorname{tr}_{e}\left[\mathrm{e}^{\mathrm{i}\hat{H}t}\delta \hat{F}_{\alpha} \mathrm{e}^{-\mathrm{i}\hat{H}t / \hbar}\left(\delta \hat{F}_{\nu} \hat{\rho}_\text{ss}+\hat{\rho}_\text{ss} \delta \hat{F}_{\nu}\right)\right].
\end{equation}

Initially, the calculation of electronic friction predominantly focuses on systems with non-interacting electrons\cite{PhysRevB.96.104305,PhysRevB.97.064303,10.1063/5.0187646}. For such systems, evaluating the electronic friction is relatively straightforward. In contrast, incorporating el-el interactions presents a significant challenge \cite{PhysRevB.88.045137,https://doi.org/10.1002/smll.200600101,PhysRevLett.94.206804,Kisiel2011,Langer2014,PhysRevB.60.5969,Kennes2017}. Nevertheless, progress has been made in determining electronic friction in the presence of el-el interactions, revealing novel and intriguing physical phenomena at low temperatures\cite{PhysRevB.58.2191,PhysRevB.60.5969,PhysRevLett.119.046001}. For instance, Ref. \cite{PhysRevLett.119.046001} employed the numerical renormalization group (NRG) method\cite{RevModPhys.47.773,RevModPhys.80.395} to study the Anderson model, though the electronic bath was still treated as non-interacting. In our previous work\cite{liu2025electronic}, we evaluated the electronic friction for interacting electrons by taking the one-dimensional Hubbard–Holstein (HH) model as an example, simulated via the density matrix renormalization group (DMRG)\cite{PhysRevLett.69.2863,PhysRevB.48.10345,RevModPhys.77.259,SCHOLLWOCK201196}. DMRG provides highly accurate correlation functions in both time and frequency domains for strongly correlated systems, however, the inherent limitations of DMRG make it difficult to extend such calculations to two-dimensional systems. 

Dynamical mean-field theory (DMFT)\cite{PhysRevLett.62.324,PhysRevLett.69.168,RevModPhys.68.13} provides a theoretical framework for strongly correlated electron systems by mapping lattice models onto self-consistent quantum impurity models. This mapping is formally exact in the limit of infinite spatial dimensions. A key point is that DMFT exhibits insensitivity to both dimensionality and boundary conditions. Here, we employ DMFT to evaluate the electronic friction of two-dimensional (2D) HH model with periodic boundary condition (PBC). To do this, we implement the NRG method\cite{RevModPhys.47.773,RevModPhys.80.395} based on the full density-matrix algorithm\cite{PhysRevLett.95.196801,PhysRevB.74.245113,PhysRevB.74.245114,PhysRevLett.99.076402,PhysRevB.92.155129} as a quantum impurity solver into the DMFT. 

This paper is organized as follows. In Sec. \ref{sec:sec2}, we introduce the HH model—a general framework for strongly correlated systems with electron-phonon coupling—and the computational methodology. Specifically, Sec. \ref{sec:sec2A} presents the dynamical mean-field theory (DMFT), and Sec. \ref{sec:sec2B} details the full density-matrix NRG (FDM-NRG) as the impurity solver within the DMFT loop. Subsequently, in Sec. \ref{sec:sec3A}, we compare the electronic friction and potential of mean force (PMF) evaluated by MFT and DMFT; in Sec. \ref{sec:sec3B}, we study the electronic friction–Langevin dynamics (EF-LD). Finally, we conclude in Sec. \ref{sec:sec4}.

\section{\label{sec:sec2} Model and Methodology}

The Hubbard-Holstein (HH) model provides an ideal testbed for studying strongly correlated systems with electron-phonon (el-ph) coupling. To establish the importance of el-el interactions for every system and bath site, we will now calculate the electronic friction for the HH model, 
\begin{align}
\hat{H}=& \hat{H}_\mathrm{el} + H_\mathrm{osc} ,  \label{eqn-4} \\
\hat{H}_\mathrm{el}=& E(x)\sum_\sigma \hat{n}_{1\sigma} + \epsilon \sum_{i\ne1,\sigma}\hat{n}_{i\sigma} + t\sum_{\langle i,j\rangle}\sum_{\sigma}\hat{c}_{i\sigma}^\dagger \hat{c}_{j,\sigma}  \nonumber \\
&+ U\sum_i\hat{n}_{i\uparrow}\hat{n}_{i\downarrow},   \label{eqn-5} \\
H_\mathrm{osc}=& \frac{p^2}{2m} + \frac{1}{2}\omega^2x^2,   \label{eqn-6}
\end{align}
where $\hat{c}^\dagger_{i\sigma}$,$\hat{c}_{i\sigma}$ are fermionic creation and annihilation operators of spin $\sigma$ on site $i$, and $\hat{n}_{i\sigma}\equiv\hat{c}_{i\sigma}^\dagger\hat{c}_{i\sigma}$. Physically, the HH model represents an electronic impurity on 1-th site near a bath and coupled to a classical vibrating oscillator with position and momentum $x$ and $p$. The impurity can be filled with an electron of up or down spin, such that  $\sigma = \uparrow,\downarrow$ indicates spin states. The oscillator is a vibrational DoF and feels a different force depending on the occupation of the impurity. We set the on-site energy of the impurity in Eq. (\ref{eqn-5}) to be $E(x)\equiv E_d + \sqrt{2}gx$. To understand how the motion of the oscillator is perturbed by the fluctuating charge of the impurity, we will evaluate the electronic friction of Hubbard-Holstein model at different temperatures.

We calculate the electronic friction tensor using Eq. (\ref{eqn-3}) within our DMFT framework, with the FDM-NRG method incorporated as the impurity solver.

\subsection{\label{sec:sec2A} Dynamical mean-field theory}

As a nonperturbative approach to the electronic structure of strongly correlated systems, the key idea of DMFT is to map a many-body lattice problem onto a self-consistently solved quantum impurity model. While the mapping itself is formally exact, the central approximation in conventional DMFT is to assume that the lattice self-energy is local and thus neglects all nonlocal correlations. This approximation becomes exact only in the limit of a lattice with an infinite coordination number.

In the DMFT calculations, the Hubbard model 
\begin{equation}\label{eqn-Hub}
    H_\mathrm{Hub} = \epsilon_0 \sum_{i\sigma}\hat{n}_{i\sigma} + t\sum_{\langle i,j\rangle}\sum_{\sigma}\hat{c}_{i\sigma}^\dagger \hat{c}_{j,\sigma} + U\sum_i\hat{n}_{i\uparrow}\hat{n}_{i\downarrow}
\end{equation}
can be mapped onto the following Anderson impurity model:
\begin{align}\label{eqn-and}
    H_\text{And} =& \epsilon_0\sum_{\sigma}d_\sigma^\dagger d_\sigma + Un_{\uparrow}n_{\downarrow} + \sum_{k\sigma}\epsilon_k c_{k\sigma}^\dagger c_{k\sigma}   \nonumber \\
    &+ \sum_{k\sigma}V_k \left(c_{k\sigma}^\dagger d_{\sigma} + \text{h.c.} \right),
\end{align}
which describes an impurity orbital $d$ coupled to a non-interacting bath. Within the DMFT framework, the influence of the bath on the impurity dynamics is fully encoded in the hybridization function 
\begin{equation}\label{eqn-hybrid}
    \Gamma(\omega)\equiv \sum_k \frac{V_k^2}{\omega+\mathrm{i}\eta-\epsilon_k},
\end{equation}
which is determined self-consistently through the DMFT loop. The impurity self-energy $\Sigma_\text{imp}(\omega)$ is obtained from the Dyson equation: 
\begin{equation}\label{eqn-self-imp}
    \Sigma_\text{imp}\left(\omega\right) = \mathcal{G}_\text{imp}^{-1}\left(\omega\right)-G_\text{imp}^{-1}\left(\omega\right),
\end{equation}
where $\mathcal{G}_\text{imp}$ denotes the non-interacting impurity Green's function:
\begin{equation}
    \mathcal{G}_\text{imp} (\omega) = \frac{1}{\omega+\mathrm{i}\eta-\epsilon_0-\Gamma(\omega)}, 
\end{equation}
and $G_\text{imp}(\omega)$ denotes the impurity Green's function to be calculated by impurity solver. Second, the lattice Green's function is 
\begin{equation}
    G_\text{lat} (\mathbf{k},\omega) = \frac{1}{\omega+\mathrm{i}\eta-\epsilon_0-\epsilon_\mathbf{k} - \Sigma_\mathrm{lat}(\mathbf{k},\omega)}.
\end{equation}
Within the DMFT approximation, the lattice self-energy is taken to be local and equal to the impurity self-energy:
\begin{equation}\label{eqn-self-appro}
    \Sigma_\mathrm{lat}\left(\mathbf{k},\omega\right) \approx \Sigma_\text{imp}\left(\omega\right) \equiv \Sigma(\omega).
\end{equation}
Therefore, for any site $i$ in the Hubbard model, the local Green's function can be derived by 
\begin{equation}\label{eqn-DMFT-G-lat}
    G_\mathrm{lat} (\omega) = \int\mathrm{d}\epsilon \frac{\mathcal{D}(\epsilon)}{\omega+\mathrm{i}\eta-\epsilon_0-\epsilon_\mathbf{k}-\Sigma(\omega)},
\end{equation}
where $\mathcal{D}(\epsilon)$ is the density of states:
\begin{equation}
    \mathcal{D}(\epsilon)\equiv \frac{1}{N_s}\sum_{\mathbf{k}} \delta(\epsilon-\epsilon_\mathbf{k}).
\end{equation}

To simplify the self-consistent calculations within DMFT, we work on the Bethe lattice, so the density of states of non-interacting lattice can be replaced as the Bethe lattice:
\begin{equation}
    \mathcal{D}(\epsilon) = \frac{2}{\pi D^2}\sqrt{D^2-\omega^2},
\end{equation}
where $D=2\tilde{t}$ stands for the corresponding half bandwidth, $\tilde{t}=\sqrt{z}t$ is the rescaled hopping coefficients, and $z$ is the coordination number. For a two-dimensional square lattice, we have $z=4$. Generally, the DMFT self-consistent condition is the lattice Green's function $G_\mathrm{lat}(\omega)$ coincides with the impurity Green's function $G_\text{imp}$, i.e. $G_\mathrm{lat}(\omega)=G_\text{imp}(\omega)$, and  the non-interacting Green's function of impurity model (i.e. the dynamical mean field) is updated self-consistently as
\begin{equation}
    \mathcal{G}_\text{imp}(\omega)= \left(\Sigma(\omega) + G_\mathrm{lat}^{-1}(\omega)\right)^{-1}.
\end{equation}
However, on the Bethe lattice, the self-consistent condition can be represented by\cite{PhysRevB.96.085118} 
\begin{equation}\label{eqn-bethe-gamma-Glat}
    \Gamma(\omega)=\frac{D^2}{4}G_\mathrm{lat}(\omega).
\end{equation}

Notably, owing to the electron-phonon coupling, the on-site energy of the first site in our electronic Hamiltonian (Eq. (\ref{eqn-5})) is different from that of the other sites. This is equivalent to introducing an impurity at the first site, thereby breaking translational symmetry. For such systems with inhomogeneous energies $\epsilon_i$, we need to employ real-space DMFT (R-DMFT)\cite{PhysRevB.70.195342,PhysRevLett.78.3943,PhysRevB.59.2549,PhysRevB.60.7834,PhysRevB.70.241104,PhysRevLett.100.056403,PhysRevLett.101.066802,Snoek_2008,PhysRevB.100.115118} to calculate their properties. The Dyson equation in real space is represented as
\begin{equation}
    G_{ij}(\omega) = \mathcal{G}_{ij}(\omega) + \sum_{kl}\mathcal{G}_{ik}(\omega)\Sigma_{kl}(\omega)G_{lj}(\omega),
\end{equation}
where $\mathcal{G}_{ij}(\omega)$ is the non-interacting Green's function. Similarly to conventional k-space DMFT, the main approximation of R-DMFT is that the self-energies are local, which means that they are diagonal in the lattice site indices,
\begin{equation}
    \Sigma_{ij}(\omega) = \Sigma_i(\omega)\delta_{ij}.
\end{equation}
Nevertheless, they are site dependent for inhomogeneous systems. For simplicity, we employ the local density approximation\cite{Snoek_2008,PhysRevB.100.115118} (LDA), which assumes the self energy is homogeneous. For example, in our electronic Hamiltonian (Eq. (\ref{eqn-5})), we assumes the on-site energy $\epsilon_i = \epsilon$, that is, the inhomogeneous part is neglected. This is equivalent to assuming that the lattice models with different on-site energy $E(x)$ at the first site have the same hybridization function (Eq. (\ref{eqn-hybrid})) when mapped into impurity models. However, when calculating the mean force $\bar{F}_\alpha$ and electronic friction coefficients $\gamma_{\alpha\nu}$ using the equivalent impurity model, the impurity level $\epsilon_0$ in Eq. (\ref{eqn-and}) is replaced by the on-site energy of the first site $E(x)$ in Eq. (\ref{eqn-5}).

We summarize the DMFT self-consistency loop as follows: 
\begin{enumerate}[noitemsep, topsep=0pt, partopsep=0pt, parsep=0pt, itemindent=*, leftmargin=*]
    \item Start with a guess dynamical mean field $\mathcal{G}_\text{imp}(\omega)$ (or hybridization function $\Gamma(\omega)$); 
    \item Compute the impurity Green's function $G_\text{imp}(\omega)$ using an impurity solver;
    \item Compute the impurity self-energy $\Sigma_\text{imp}\left(\omega\right)$ from Eq. (\ref{eqn-self-imp}); 
    \item Make the DMFT approximation: Eq. (\ref{eqn-self-appro});
    \item Compute the lattice Green's function $G_\mathrm{lat}\left(\omega\right)$ from Eq. (\ref{eqn-DMFT-G-lat});
    \item Update the hybridization function $\Gamma(\omega)$ by Eq. (\ref{eqn-bethe-gamma-Glat});
    \item Go back to step 2 until convergence;
    \item Calculate the properties such as $\bar{F}_\alpha$ and $\gamma_{\alpha\nu}$ using equivalent hybridization function $\Gamma(\omega)$.
\end{enumerate}

\subsection{\label{sec:sec2B} Full density-matrix NRG}

The NRG procedure\cite{RevModPhys.47.773,RevModPhys.80.395} begins by discretizing the continuous hybridization function on a logarithmic energy grid defined by $\pm D\Lambda^{-n}$ ($\Lambda > 1$, $n=0,1,2,\cdots$). Through a standard tridiagonalization — which effectively retains only the lowest-energy mode within each discretization interval — the original impurity model is mapped onto a semi-infinite tight-binding chain whose hopping amplitudes decay exponentially along the chain. The first site of this chain represents the impurity. The Hamiltonian is then diagonalized iteratively by adding one site at a time. To prevent the Hilbert space from growing exponentially, at each step $N$, the eigenstates of the enlarged chain Hamiltonian $H_N$ are constructed from the states of the newly added ($N$-th) site and only the $M$ lowest-lying eigenstates (the “reserved states”) of the previous Hamiltonian $H_{N-1}$; all higher-energy states from $H_{N-1}$ are discarded. This iterative process continues until the hopping between the last added site ($N = N_\text{max}$) and its neighbor becomes the smallest relevant energy scale, at which point the full-chain Hamiltonian $H_{N_\text{max}}$ provides a faithful low-energy representation of the original Anderson impurity model.

The full density-matrix NRG (FDM-NRG)\cite{PhysRevLett.99.076402} method represents a fundamental extension of the traditional numerical renormalization group (NRG) approach. A complete basis set\cite{PhysRevLett.95.196801,PhysRevB.74.245113,PhysRevB.74.245114} introduced in FDM-NRG enables the evaluation of the full density matrix at finite temperature, thereby allowing for accurate calculations of both thermodynamic quantities\cite{PhysRevB.92.155129} and, most importantly, dynamical correlation functions\cite{PhysRevB.74.245114}. The key advancement lies in its ability to produce significantly more reliable spectral functions, with improved resolution of fine structures such as Kondo resonances, side bands, and high-energy features, which are often broadened or lost in the standard NRG due to its truncation of the state space.

The complete basis set\cite{PhysRevLett.95.196801,PhysRevB.74.245113,PhysRevB.74.245114} of the Fock space of $H_{N_\text{max}}$ is used to constructed the tensor-product state $\left|se\right\rangle_N\equiv \left|s\right\rangle_N\otimes \left|\alpha_{N+1}\right\rangle \otimes\left|\alpha_{N+2}\right\rangle \cdots \otimes \left|\alpha_{N_\text{max}}\right\rangle$ from $\left|s\right\rangle_N$ the $s$-th discarded states of $H_N$ and $\left|\alpha_m\right\rangle$ the state of the $m$-th site with $\left\{\left|0\right\rangle, \left|\uparrow\right\rangle, \left|\downarrow\right\rangle, \left|2\right\rangle\right\}$, where $e$ denotes collectively the degrees of freedom of the sites $m=N+1, \cdots,N_\text{max}$, i.e., the environment of $H_N$. The completeness relation of the basis set $\left|se\right\rangle_N$ reads as
\begin{equation}
    1 = \sum_{N=n_0}^{N_\text{max}} \sum_{se}\left|se\right\rangle_N^\text{D} {}_N^\text{D}\left\langle se\right|,
\end{equation}
along with the orthonormality
\begin{equation}
    {}_{N}^\text{D}\left\langle se | s^\prime e^\prime \right\rangle_{N^\prime}^\text{D}=\delta_{NN^\prime}\delta_{ss^\prime}\delta_{ee^\prime},
\end{equation}
where $n_0$ is the first iteration at which high-energy states are discarded, "D" denotes the discarded states. Note that we take all eigenstates of the last iteration $N_\text{max}$ as discarded.

At each iteration $N$, the Fock space $\mathcal{F}_{N_\text{max}}$ of a Wilson chain with fixed length $N_\text{max}$ is partitioned by all previously discarded states
\begin{equation}
    1_N^- = \sum_{N^\prime=n_0}^{N-1} \sum_{se} \left|se\right\rangle_{N^\prime}^\text{D} {}_{N^\prime}^\text{D}\left\langle se\right|,
\end{equation}
and all states at iteration $N$
\begin{align}
    1_N^+ =& \sum_{N^\prime=N}^{N_\text{max}}\sum_{se}\sum_{se} \left|se\right\rangle_{N^\prime}^\text{D} {}_{N^\prime}^\text{D}\left\langle se\right|  \nonumber \\
    =& \sum_{se} \left|se\right\rangle_N^\text{X} {}_N^\text{X}\left\langle se\right|,
\end{align}
i.e., $1=1_N^-+1_N^+$. Here, $\text{X}=\text{R},\text{D}$, and $\text{R}$ represent the reserved states. Since $\left|se\right\rangle_N$ is only an exact eigenstate of $H_N$ corresponding to an eigenvalues $E_s^N$ with $d^{N_\text{max}-N}$-fold degeneracy, where $d$ is the degree of freedom of a single bath site, one has to assume it is also an eigenstate of the original model $H\left|se\right\rangle_N\approx E_s^N\left|se\right\rangle_N$. This is the only approximation of the FDM algorithm. The full density matrix $\rho$ of $H$ is represented as\cite{PhysRevLett.99.076402} 
\begin{align}
    \rho =& \frac{1}{Z}\sum_{N=n_0}^{N_\text{max}}\sum_{se}\mathrm{e}^{-\beta E_s^N}\left|se\right\rangle_N^\text{D}{}_N^\text{D}\left\langle se\right|=\sum_N\omega_N\rho_\text{DD}^N,  \\
    \omega_N =& d^{N_\text{max}-N}Z_N^\text{D}/Z,  \\
    \rho_\text{DD}^N =& \sum_{s}^\text{D}\frac{\mathrm{e}^{-\beta E_s^N}}{Z_N^\text{D}}\left|se\right\rangle_N^\text{D}{}_N^\text{D}\left\langle se\right|,  \\
    Z_N^\text{D} =& \sum_s^\text{D}\mathrm{e}^{-\beta E_s^N},  \\
    Z =& \sum_N\sum_s^\text{D}d^{N_\text{max}-N}\mathrm{e}^{-\beta E_s^N}=\sum_N\sum_s^\text{D}d^{N_\text{max}-N}Z_N^\text{D},
\end{align}
with $\beta\equiv 1/\left(k_\text{B}T\right)$ being the inverse temperature. Using this form of the density matrix, the complete base set of discarded states, and the NRG approximation, all the dynamic and static properties of Hamiltonian $H$ can be evaluated.

\section{\label{sec:sec3} Results and Discussions}

\subsection{\label{sec:sec3A} Electronic friction and potential of mean force}

\begin{figure*}[htbp]
    \centering
    \includegraphics[width=0.8\linewidth]{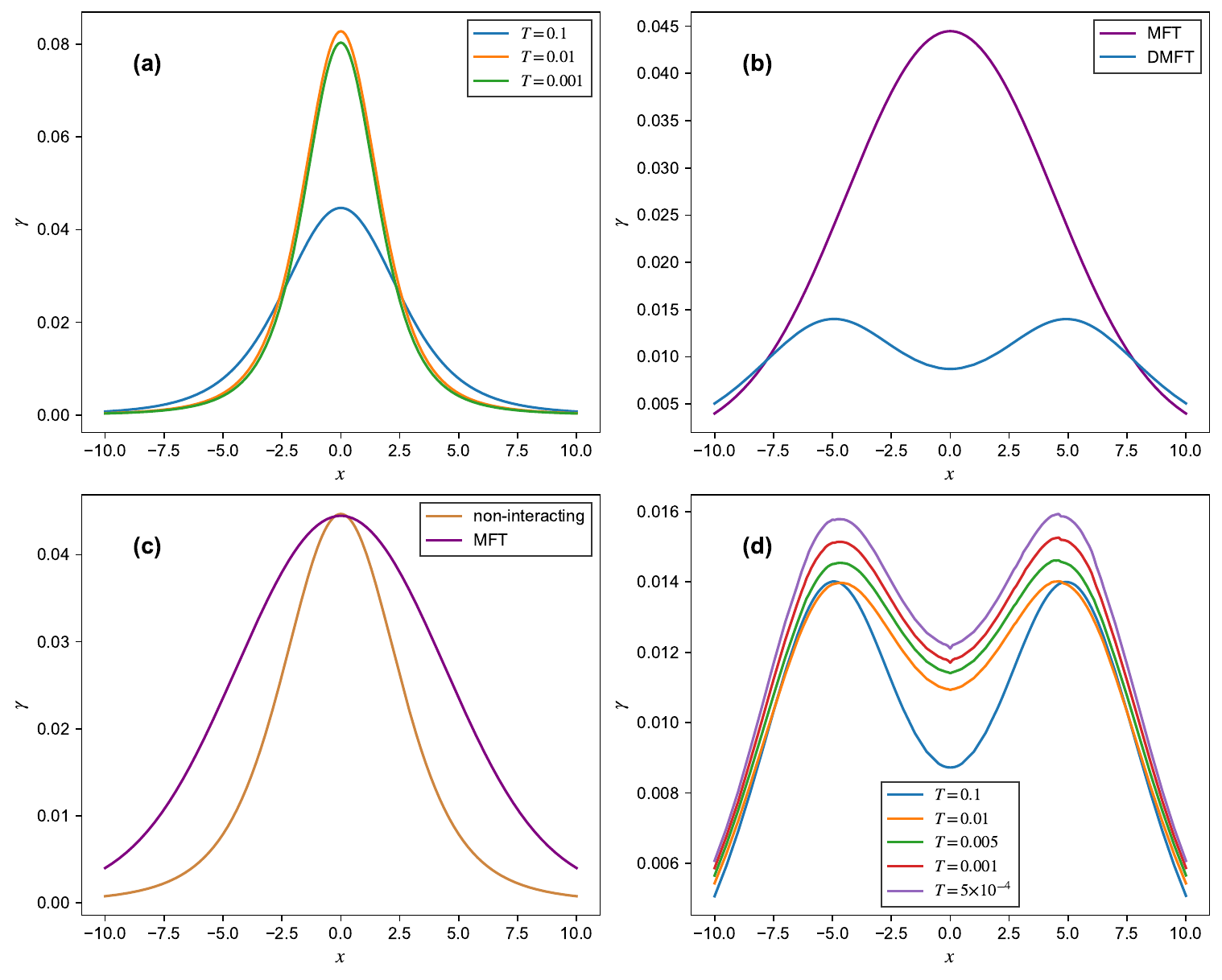}
    \caption{\label{fig-1}{(a) Electronic friction of the non-interacting lattice model ($U=0$) as a function of position $x$. Only a single peak occurs in the electronic friction. (b) Electronic friction obtained from MFT and DMFT calculations at temperature $T=0.1$. Due to electron-electron repulsions in the Hubbard model, DMFT successfully captures two peaks in the friction, whereas MFT fails to reproduce them. (c) Electronic friction of the non-interacting lattice model and the Hubbard-Holstein model obtained from MFT at temperature $T=0.1$. Within MFT, the friction merely corresponds to a broadening of peak without el-el interactions, indicating that it still retains many single-particle features. (d) Electronic friction as a function of position $x$ according to DMFT. For systems without el-el interactions, we set $\epsilon=U=0$; for systems with el-el interactions, we set $\epsilon=-0.5$ and $U=1$. The remaining parameters are fixed as $E_d = -0.5$, $g = 0.075$, $t = 0.05$, and we set $k_B = \hbar = 1$.}}
\end{figure*}

We employ DMFT using FDM-NRG as an impurity solver to compute the electronic friction as a function of $x$ for the two-dimensional HH model (Eqs. (\ref{eqn-4})-(\ref{eqn-6})); the results are shown in Fig. \ref{fig-1}. In Fig. \ref{fig-1}(a), we show that the electronic friction in the absence of el-el interactions ($U=0$). A single peak appears, corresponding to the resonance condition where electron attachment or detachment aligns with the Fermi level of the solid surface $\epsilon_F$, i.e. $E_d + \sqrt{2}gx=0$ (we have set $\epsilon_F=0$). As the temperature decreases, the peak height increases. Reducing the temperature further from $T=0.01$ to $T=0.001$, leaves the friction almost unchanged, indicating that we have reached the zero-temperature limit. In Fig. \ref{fig-1}(b), we compare the electronic friction calculated from DMFT versus the results from MFT at temperature $T=0.1$. Notice that DMFT predicts two peaks in the electronic friction, indicating the existence of a new energy level due to el-el interactions. Therefore, resonances of electron attachment or detachment in impurity with the Fermi level of other sites within the Hubbard model occur near $E_d + \sqrt{2}gx=0$ and $E_d + \sqrt{2}gx+U=0$. In contrast, MFT only predicts one peak at the position in the middle of two Fermi resonance, where DMFT predicts a dip. Moreover, the peaks from DMFT are lower than those from MFT. In Fig. \ref{fig-1}(c), we observe that within MFT the friction merely corresponds to a broadening of the peak found in the non-interacting case, indicating that MFT still retains many single-particle features. Finally, Fig. \ref{fig-1}(d) shows the electronic friction of the two-dimensional Hubbard-Holstein model at different temperatures, calculated using DMFT. As the temperature decreases, the electronic friction effects become more pronounced.

\begin{figure*}[htbp]
    \centering
    \includegraphics[width=0.75\linewidth]{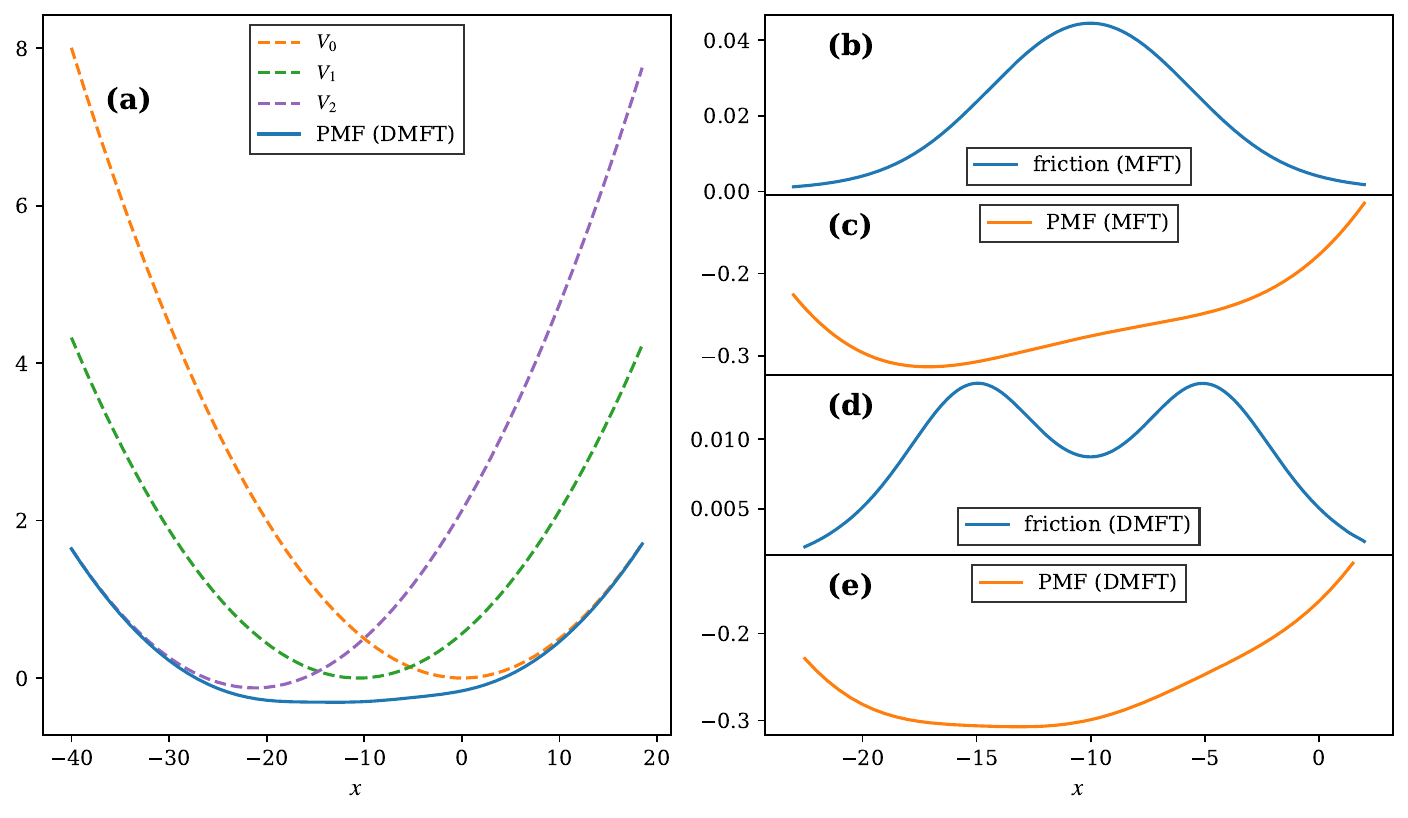}
    \caption{\label{fig-2}{(a) PMF and diabatic PESs as a function of position $x$. (b) and (c) Electronic friction and PMF as a function of position $x$ according to MFT, respectively. (d) and (e) Corresponding results obtained from DMFT. The parameters are $g = 0.075$, $\omega = 0.1$, $t = 0.25$, $U = 1$, $E_d=\frac{g^2}{m\omega^2}$, $T=0.1$, and we set $m = k_B = \hbar = 1$.}}
\end{figure*}

In the diabatic picture, there are three different potential energy surfaces (PESs)-those with the impurity unoccupied (denoted as 0), those with the impurity occupied by only one electron (denoted as 1), and those with the impurity occupied by two electrons (denoted as 2):
\begin{align}
H_{\alpha} & =\frac{p^2}{2m} + V_{\alpha}, \alpha=0,1,2 \\
V_{0} & =\frac{1}{2} \omega^2 x^{2}, \\
V_{1} & =\frac{1}{2} \omega^2 x^{2} + E(x), \\
V_{2} & =\frac{1}{2} \omega^2 x^{2} + 2E(x) + U.
\end{align}
Also, the potential of mean force (PMF) is defined as\cite{Bode2012}
\begin{equation}\label{eqn-29}
V_\text{PMF}=\frac{1}{2}\omega^2 x^2 - \int_{x_0}^x\mathrm{d}x^\prime \bar{F}_x\left(x^\prime\right),
\end{equation}
where $\bar{F}_x$ is the mean force along $x$ direction defined as Eq. (\ref{eqn-1}).

In Fig. \ref{fig-2}(a), we plot the PMF calculated by DMFT and PESs as a function of $x$ for the HH model. Unlike the one-dimensional system calculated using DMRG, the PMF does not exhibit pronounced peaks at the intersections of the PESs $V_0$ and $V_1$, as well as $V_1$ and $V_2$. However, results from the one-dimensional system show that both the PMF and the electronic friction coefficients reach maximum values at the intersections of the PESs\cite{liu2025electronic}. Near these positions, partial occupancy of the impurity electron promotes electron exchange between the molecule and the solid surface. Regarding the absence of clear peaks in the PMF, we propose two possible explanations. On the one hand, introducing el-el interactions in the two-dimensional system significantly suppresses electron exchange at the PES intersections. As can be seen from the calculated electronic friction coefficients, the peaks are considerably smaller when el-el interactions are present than in the non-interacting case, providing strong evidence for the above argument. On the other hand, DMFT, as an approximate method (whose reliability is at least lower than that of DMRG), still requires further validation of its computational accuracy. Figures (b) and (c) show the electronic friction coefficients and the PMF calculated using MFT, respectively, while Figures (d) and (e) present the corresponding results obtained from DMFT. These results are prerequisites for subsequent electronic friction–Langevin dynamics (EF-LD) simulations. Note that, here we set $E_d=\frac{g^2}{m\omega^2}$ to conveniently demonstrate the relationship between PESs, PMF and electronic friction.

\subsection{\label{sec:sec3B} Electronic friction-Langevin dynamics(EF-LD)}

\begin{figure*}[htbp]
    \centering
    \includegraphics[width=0.75\linewidth]{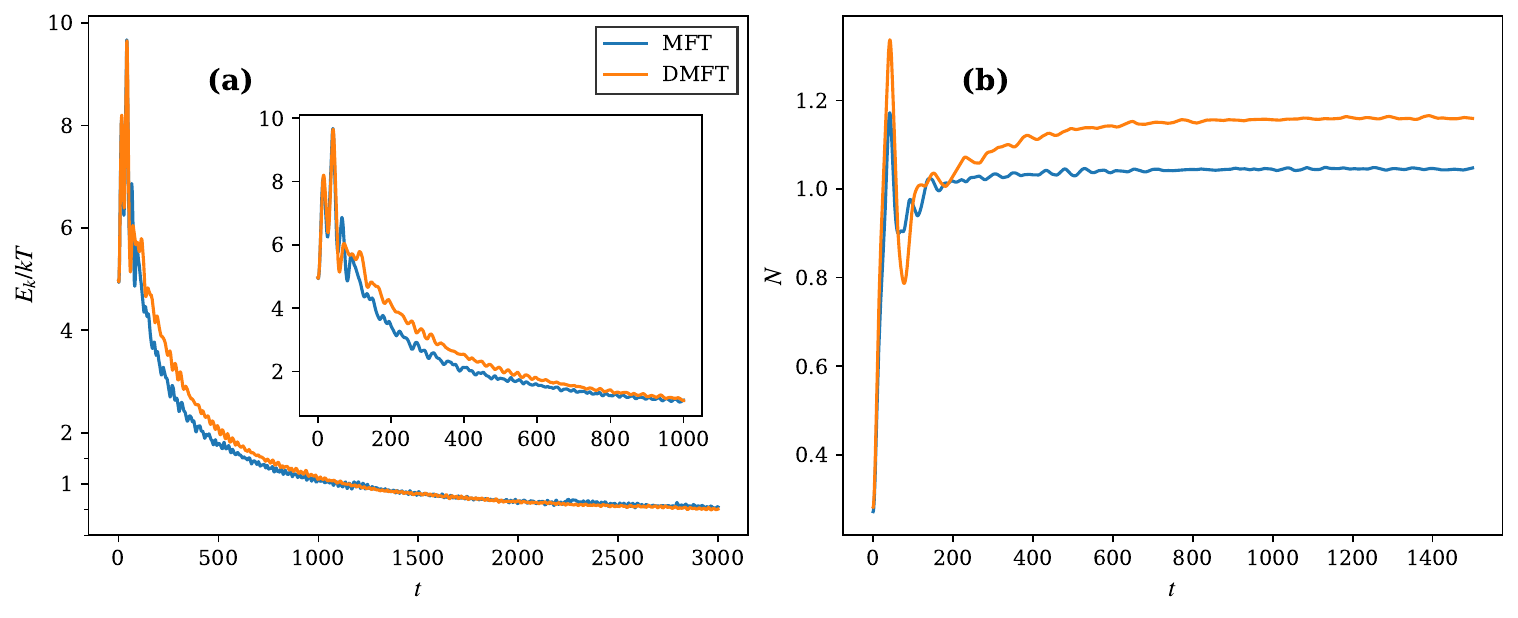}
    \caption{\label{fig-3}{The results from EF-LD. (a) average kinetic energy; (b) electronic population in the impurity. The parameters are $g = 0.075$, $\omega = 0.1$, $t = 0.25$, $U = 1$, $E_d=\frac{g^2}{m\omega^2}$, $T=0.1$, and we set $m = k_B = \hbar = 1$. We initialize the oscillators at $x=0$ by sampling their states from a Boltzmann distribution, where the average initial kinetic energy per oscillator is $5k_\mathrm{B}T$.}}
\end{figure*}

We will now study the nonadiabatic dynamics within the electronic frictional model, to compare the EF-LD of two-dimensional HH model based on MFT- and DMFT-based mean force and electronic friction.

Although the nonadiabatic dynamics of a molecule interacting with a two-dimensional solid surface is studied here, the molecule is assumed to move along one-dimensional coordinate. Therefore, the one-dimensional Langevin equation (\ref{eqn-1}) can still be used in the EF-LD simulations, which can be simplified as follows,
\begin{align}
m \dot{v}_x&=\bar{F}_{x} -\gamma{v}_{x}+\zeta(t),  \label{eqn-1D-Langevin} \\
v_x&=\frac{\mathrm{d}x}{\mathrm{d}t},   \label{eqn-velocity} 
\end{align}
where the random force $\zeta\left(t\right)$ is assumed to be a Gaussian variable with a norm $\sigma = \sqrt{\frac{2\gamma mkT}{\mathrm{d} t}}$. This condition satisfies the second fluctuation-dissipation theorem\cite{10.1063/1.4733675}. $\mathrm{d} t$ is the time step interval. We then use 4th order Runge-Kutta (RK4) to integrate Eqs. (\ref{eqn-1D-Langevin}) and (\ref{eqn-velocity}), where 10000 trajectories have been used for the EF-LD simulations. We initialize the oscillators at $x=0$ by sampling their states from a Boltzmann distribution, where the average initial kinetic energy per oscillator is $5k_\mathrm{B}T$. The random force $\zeta(t)$ is generated by a normal distribution.    

The average kinetic energy and impurity electronic population obtained from EF-LD simulations are shown in Fig. \ref{fig-3}. These simulations are performed using the electronic friction coefficients and PMF derived from both MFT and DMFT. According to the equipartition theorem, the long-time limit of the average kinetic energy should be $\frac{1}{2}kT$, and indeed both MFT and DMFT results converge to this value. The nuclear dynamics predicted by the two methods are also in good mutual agreement. In contrast, their electronic dynamics show pronounced differences: at short times, the electronic population from DMFT deviates from that of MFT; at longer times, the equilibrium populations also differ significantly between the two approaches. These discrepancies can be traced back to Fig. \ref{fig-2}, which highlights clear differences in the mean forces and electronic friction coefficients entering the Langevin equation (Eq. (\ref{eqn-1D-Langevin})). Hence, the dynamical outcomes from DMFT and MFT are expected to be substantially different.

\section{\label{sec:sec4} Summary}

In this work, we have systematically evaluated the electronic friction in the two-dimensional Hubbard–Holstein model as a function of the impurity position $x$. Our approach combines dynamical mean-field theory (DMFT) with the full density-matrix numerical renormalization group (FDM-NRG) as the impurity solver within the DMFT self-consistent loop. Within this framework, DMFT predicts two distinct peaks in the electronic friction, corresponding to electron attachment and detachment resonances at the Fermi level of the solid surface, a direct consequence of electron–electron interactions, whereas MFT still struggles to yield qualitatively consistent results for two-dimensional systems.

We then present the PMFs obtained from MFT and DMFT, and further perform EF-LD simulations based on these PMF and electronic friction. The peaks of the PMF in the two-dimensional system calculated using DMFT are less pronounced than those in the one-dimensional system calculated with DMRG\cite{liu2025electronic}. This may be because electron–electron interactions in the two-dimensional system significantly suppress electron exchange at the crossings of potential energy surfaces, or because the computational accuracy of DMFT is still insufficient. Although the average kinetic energies from EF-LD calculations based on MFT and DMFT agree well, the deviations in the computed electronic friction and PMF within the MFT framework are reflected in the electron occupancy dynamics: the EF-LD results based on MFT deviate significantly from the more reliable DMFT-based results. This demonstrates the importance of going beyond MFT for simulating the dynamical processes of strongly correlated systems.

DMFT is better suited for two-dimensional systems compared with DMRG. Meanwhile, DMFT has already been widely applied to realistic systems, thus EF-LD studies based on DMFT can be extended to realistic systems for the \textit{ab initio} calculations. Furthermore, the strategy developed here can be extended to to nonequilibrium scenarios using nonequilibrium DMFT\cite{RevModPhys.86.779,PhysRevB.78.235124}. However, although the DMFT results are at least qualitatively consistent based on physical intuition and by comparison with one-dimensional calculations, as an approximate method its validity still requires further cross-validation.

\begin{acknowledgments}
W. D. thanks the funding from National Natural Science Foundation of China (No. 22273075 and No. 22361142829) and Zhejiang Provincial Natural Science Foundation (No. XHD24B0301). Y. L. thanks Zhecun Shi and Prof. Wei Zhu for instrumental guidance on DMFT. 
\end{acknowledgments}


\bibliography{apssamp}

\end{document}